\documentclass[showpacs,twocolumn,superscriptaddress]{revtex4}

\usepackage{amsmath}
\usepackage{amssymb}
\usepackage{bm}      
\usepackage{graphicx}

\begin{document}
\title{Quantum information processing via a lossy bus}

\author{S. D.\ Barrett} \email{seandbarrett@gmail.com}
\affiliation{Hewlett-Packard Laboratories, Filton Road Stoke Gifford,
  Bristol BS34 8QZ, United Kingdom}
\affiliation{Centre for Quantum Computer Technology, University of
Queensland, St Lucia, Queensland 4072, Australia}
\affiliation{Blackett Laboratory, Imperial College London, Prince
Consort Road, London SW7 2BW, United Kingdom}

\author{G. J. Milburn}
\affiliation{Centre for Quantum Computer Technology, University of
Queensland, St Lucia, Queensland 4072, Australia}

\date{\today}

\begin{abstract}
We describe a method to perform two qubit measurements and logic
operations on pairs of qubits which each interact with a harmonic
oscillator degree of freedom (the \emph{bus}), but do not directly
interact with one another. Our scheme uses only weak interactions
between the qubit and the bus, homodyne measurements, and single
qubit operations. In contrast to earlier schemes, the technique
presented here is extremely robust to photon loss in the bus mode,
and can function with high fidelity even when the rate of photon
loss is comparable to the strength of the qubit-bus coupling.
\end{abstract}



\maketitle

The practical implementation of quantum information processing (QIP)
devices is one of the central aims of quantum information science.
Large scale quantum computers can dramatically outperform classical
computers for certain problems \cite{Shor94}, while few qubit
devices would allow other useful QIP applications, including quantum
repeaters \cite{Briegel1998} and cryptography \cite{Gisin2002}.

Of particular interest are \emph{bus mediated} QIP architectures, in
which coherent quantum operations on multiple qubits are implemented
indirectly, via a continuous degree of freedom (the \emph{bus}) such
as a quantized electromagnetic field mode which interacts with each
of the qubits. Such architectures have a number of potential
advantages for practical implementations. Firstly, in some systems,
engineering a coherent interaction between qubits and a bus mode can
be easier than directly coupling pairs of qubits. Secondly,
architectures involving a bus mode can be more readily scaled to
large numbers of qubits. When a new qubit is added to the system,
only it's interaction with the bus mode needs to be calibrated and
controlled. Bus mediated operations can then be implemented on
\emph{any} pair of qubits, rather than being restricted to (say)
nearest neighbour qubits. Finally, in many cases the bus can be
propagated long distances, which can be useful for distributed
applications such as quantum repeaters.

Recently, additional interest in bus-mediated QIP has been generated
by several experiments which have demonstrated the coherent coupling
between qubits and continuous degrees of freedom. These experiments
include the strong coupling of a superconducting charge qubits to a
single electromagnetic field mode \cite{Walraff2004}, the coupling
of trapped ion qubits to a collective vibrational mode
\cite{Leibfried2003} and the observation of strong coupling in
various quantum dot-cavity and atom-cavity systems
\cite{Reithmaier2004,Yoshie2004,Boca2004,Keller2004}.

Previously, one of us introduced a scheme for implementing two-qubit
operations via a bus mode \cite{Barrett2005}. This was described in
terms of an all-optical implementation, using a weak cross-Kerr
interaction between the photonic qubits and a bus which is a single
mode of the optical field, though the scheme could be modified to
work for other systems with a suitable qubit-bus coupling
Hamiltonian. In this scheme, the bus mode is initially prepared in a
coherent state $|\alpha \rangle = e^{-|\alpha|^2/2} \sum_n \alpha^n
|n \rangle /\sqrt{n!}$, where $|n\rangle$ denotes the $n$'th excited
state of the oscillator. Subsequently the bus interacts with the
qubit degrees of freedom via appropriately chosen cross-Kerr
interactions. Finally, by performing a homodyne measurement on the
bus degree of freedom, the bus is disentangled from the qubits, and
the entanglement is transferred to the qubits. In the original
proposal \cite{Barrett2005}, this technique was used to implement
two-qubit entangling measurements on the qubits, which are
sufficient for universal quantum computation \cite{Rudolph2005}. A
modification of the scheme also allows the implementation of a CNOT
gate \cite{Nemoto2004}. Other schemes for implementing gates using a
bus mode have also been proposed \cite{Spiller2006,Beige2006}.

However, these schemes have a substantial drawback with regard to
practical implementations. At various stages in the protocol, the
bus mode is in a superposition of coherent states, $|\alpha e^{i
\theta_q}\rangle$, with distinct phases $\{ \theta_q \}$ which
correspond to different states of the qubits. Superpositions of this
form are rather fragile in the presence of dissipation (e.g. photon
loss) of the bus degree of freedom. Hence, losing even a small
number of quanta from the bus can lead to complete decoherence of
the combined qubit-bus state, and destroy any entanglement between
the qubits. A careful analysis shows that the scheme introduced in
\cite{Barrett2005}, and those derived from it \cite{Nemoto2004}, can
only work in the regime of extremely low dissipation defined by
$\kappa / \chi \ll \chi \tau $, where $\kappa$ is the loss rate for
quanta in the bus mode, $\chi$ is the interaction strength between
the qubits and the bus, and $\tau$ is the corresponding interaction
time. Since, in most realistic implementations, the bus mode will be
subject to dissipation, this requirement can be a significant
restriction.

In this Letter, we describe a method for bus mediated quantum
information processing which is \emph{robust to dissipation} of the
bus mode. The physical resources used for this scheme are similar to
those used in the original scheme \cite{Barrett2005}, namely
preparation of a coherent state of the bus mode, conditional phase
shifts of the bus and homodyne measurements. In addition, the new
scheme also makes use of unconditional \emph{displacements} of the
bus mode; such displacements are straightforward to implement in
most systems, e.g. by driving the bus mode with a classical field
resonant with the oscillator frequency. In particular, we show how,
with these resources, a near-deterministic, two qubit projective
parity measurement (i.e. a measurement of the operator $Z_1 Z_2$)
can be implemented. Such a measurement, when augmented with single
qubit operations, is universal for quantum computation, either using
the CNOT construction of Ref. \cite{Pittman2001} or by constructing
cluster states \cite{BrowneRudolph2005,BarrettKok2005}. The
interaction between the qubits and the bus can be controlled in such
a way that the scheme is robust to losses in the bus mode even in
the regime $\kappa > \chi$, where the strength of the qubit-bus
interaction is weak compared the loss rate.

\begin{figure}[!hbt]
\begin{center}
\includegraphics[scale=1.0]{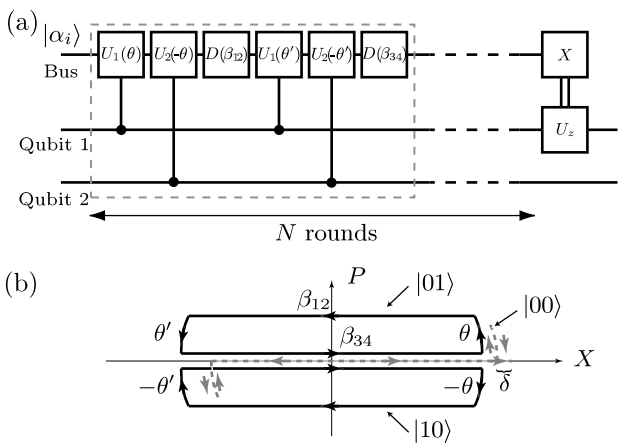}
\end{center}
\caption{The photon loss tolerant scheme. (a) The `circuit' that
implements the scheme. (b) A phase space illustration of a single
round of the protocol. In the absence of dissipation, we would have
$\theta' = -\theta$, $\beta_{12}=-2\alpha_i \cos \theta$ and
$\beta_{34}= 2\alpha_i$. Dissipation leads to a slight deformation
of the bus' trajectory.} \label{figPhaseSpaceCartoon}
\end{figure}

Our scheme for implementing parity measurements is illustrated in
Fig. \ref{figPhaseSpaceCartoon}. The bus is prepared in the coherent
state $|\alpha_i\rangle$, with $\alpha_i$ real. The bus then
interacts with the qubits via repeated applications of the circuit
shown in Fig. \ref{figPhaseSpaceCartoon}(a). The $U_j(\theta)$
operation is a conditional phase shift operation, which acts on
qubit $j$ and the bus mode as $|0\rangle |\alpha\rangle \to
|0\rangle |\alpha e^{i \theta}\rangle$ and $|1\rangle |\alpha\rangle
\to |1\rangle |\alpha e^{-i \theta}\rangle$ (throughout this Letter,
the states are written with the ket for the bus mode at the right).
$D(\beta) = e^{\beta a^\dag - \beta^* a}$ is the displacement
operator acting on the bus mode. The effect of a single round of the
protocol can be understood by considering the phase space diagram in
Fig. \ref{figPhaseSpaceCartoon}(b). For the 2-qubit input states
$|01\rangle$ and $|10\rangle$, the bus mode follows the indicated
loops in phase space. $\theta$, $\theta'$, $\beta_{12}$ and
$\beta_{34}$ are chosen such that the bus returns its initial state,
$|\alpha_i\rangle$, at the end of each round of the protocol. For
the input states $|00\rangle$ and $|11\rangle$, the consecutive
$U_j(\theta)$ operations cancel each other out, and the bus mode
follows a path along the $x-$axis. Since $\beta_{34} > \beta_{12}$,
the final state of the bus, for the even parity states, is displaced
by a small distance $\delta \approx \alpha_i \theta^2$. By repeating
this procedure $N$ times, the relative displacement of the bus mode
for the odd and even parity qubit inputs can be made sufficiently
large that the two cases can be distinguished by a homodyne
measurement of the $x-$quadrature of the bus mode. Finally, if the
measurement outcome indicates an odd parity state, a fixed rotation
about the $z-$axis is applied to one of the qubits to correct for an
unwanted phase accumulation during the protocol.

As we show below, by making $\theta$ sufficiently small, this
procedure can be made extremely tolerant to photon loss.
Heuristically, this works because, for a given input state parity,
the bus mode is never in a superposition of coherent states with
substantially different complex amplitudes. Thus, while a moderate
amount of photon loss reduces the total amplitude of the bus state,
it does not destroy the coherence of the superposed states within
the relevant parity subspaces.

To analyze this scheme in more detail, we consider a pair of qubits
coupled to a harmonic oscillator (the bus) via the interaction
picture Hamiltonian ($\hbar = 1$)
\begin{equation}
H_I(t) = \frac{\chi_1(t)}{2} Z_{1} a^\dag a + \frac{\chi_2(t)}{2}
Z_{2} a^\dag a + i \varepsilon(t) (a^\dag - a)\,. \label{eqH}
\end{equation}
Here, $\chi_i(t)$ denotes the strength of the coupling between qubit
$i$ and the bus mode, $Z_{i}$ is the Pauli $z-$operator for the
$i$th qubit, $a$ is a lowering operator for the bus, and
$\varepsilon(t)$ denotes the strength of the external driving field
acting on the bus, which is resonant with the bare oscillator
frequency \footnote{Note Eq. (\ref{eqH}) is defined in an
interaction picture relative to the bare Hamiltonian of the
uncoupled qubits and bus, $H_0 = H_{q1} + H_{q2} +
H_{\textrm{bus}}$, where $H_{qi} = -\frac{E_i}{2} Z_{i}$ and
$H_{\textrm{bus}} = \omega_0 a^\dag a$, with $E_i$ the qubit energy
splittings, and $\omega_0$ the bare oscillator frequency. In
arriving at Eq. (\ref{eqH}), we have made a rotating wave
approximation for the driving term, which is valid provided
$\omega_0 \gg \varepsilon(t)$, and provided $\varepsilon(t)$ changes
slowly on timescales of order $\omega_0^{-1}$.}. We assume the phase
and amplitude of this driving term can be accurately controlled. The
first two terms in Eq. (\ref{eqH}) generate the conditional phase
shifts of the oscillator, while the third term generates the
unconditional displacements of the oscillator. This Hamiltonian can
be realized in a variety of settings, including the dispersive
regime of cavity quantum electrodynamics (CQED) \footnote{In a CQED
implementation, additional multi-qubit terms can appear in the
Hamiltonian due to higher order interactions via the bus. These can
be avoided by engineering the system parameters such that
$\mathrm{sgn}(\Delta_1) = -\mathrm{sgn}(\Delta_2)$, $g_1 \ne g_2$,
and $g_{1,2} \lesssim |\Delta_1+\Delta_2|$ where $g_i$ are the
qubit-cavity coupling strengths, and $\Delta_i$ are the qubit-cavity
detunings.} (which can be realized in a variety of atomic and
solid-state systems, and can also be simulated in an ion trap system
\cite{Leibfried2003}), and in an all-optical setting, via the
cross-Kerr effect \cite{Barrett2005}.

A single round of the scheme can be implemented by a piecewise
constant time dependence for $\chi_i(t)$ and $\varepsilon(t)$. We
take $\chi_1(t) = -\chi_2(t) = \{ \chi_0,\,0,\,  -\chi_0,\, 0\}$,
and $\varepsilon(t) = \{0,\, -\varepsilon_0,\, 0,\, \varepsilon_0\}$
for the intervals $\{ 0<t<t_1,\, t_1<t<t_2,\, t_2<t<t_3,\,
t_3<t<t_4\}$. This has the effect of implementing the consecutive
pairs of $U_j(\pm\theta)$ operations [shown in Fig.
\ref{figPhaseSpaceCartoon}(a)] simultaneously, which is conceptually
equivalent to performing them consecutively, since the Hamiltonian
terms that generate these operations commute with one another. We
also define the time intervals $\tau_i = t_i-t_{i-1}$ corresponding
to the length of each step of the evolution. The results obtained
using this choice for $\chi_i(t)$ and $\varepsilon(t)$ will be
qualitatively valid for other choices, too. For example, in a
particular implementation, it may be easier to have a fixed,
constant qubit-bus coupling, and to periodically apply very strong,
short driving pulses (with $|\varepsilon| \gg |\chi_i|$) to the bus
mode.

We describe dissipation using a standard approach for the CQED
setting: the damping from the oscillator may be modeled as leakage
from the corresponding cavity mirror at a rate $2\kappa$. Within the
Born, Markov, and rotating wave approximations, such losses can be
described by the non-unitary term in a Lindblad master equation
\cite{WallsMilburn}:
\begin{equation}
\dot{\rho}(t) = - i \left[H_I(t),\rho(t)\right] + 2 \kappa
\mathcal{D}[a] \rho(t), \label{eqFullMasterEquation}
\end{equation}
where $\rho(t)$ is the density matrix of the combined system of both
qubits and the oscillator mode, and $\mathcal{D}[A] \rho = A \rho
A^\dag - (\rho A^\dag A + A^\dag A \rho)/2$ \footnote{Equation
(\ref{eqFullMasterEquation}) describes errors only due to damping of
the oscillator mode; in general, there will also be decoherence
processes acting directly on the qubit degrees of freedom. However,
such processes are typically independent of those due to damping and
can be dealt with using standard fault tolerance techniques, and so
we omit them from this analysis.}. Below, we solve Eq.
(\ref{eqFullMasterEquation}) for a single round of the protocol, and
then use the results to determine both the total number of rounds
required to distinguish even parity states from odd parity states,
and the total dephasing error at the end of these $N$ rounds.

The solution of Eq.(\ref{eqFullMasterEquation}) can be simplified by
noting that, since the aim of the scheme is to perform a parity
measurement, only dephasing errors \emph{within} the corresponding
parity subspaces are of concern. Thus we are only interested in
solutions of Eq. (\ref{eqFullMasterEquation}) for initial states of
the form $(x|00\rangle + y|11\rangle)  | \alpha_i \rangle$ and
$(x|10\rangle + y|01\rangle)  | \alpha_i \rangle$, where $x$ and $y$
are arbitrary amplitudes. We take the worst case, $x = y =
1/\sqrt{2}$, and so the resulting errors should be viewed as an
upper bound. The solution can be further simplified by projecting
$\rho(t)$ onto the computational basis of the qubits. This results
in a set of uncoupled differential equations for the operators
$\rho_{i|j}(t) = \langle i | \rho(t) | j \rangle$, where $i,j \in
\{00,01,10,11 \}$, which can be solved independently. These
equations are analogous to those describing the single-qubit, driven
Jaynes-Cummings model in the dispersive limit, and thus can be
solved exactly using techniques similar to those used in Refs.
\cite{PiexotodeFaria1999,PiexotodeFaria2004}. We omit the details of
this calculation, but give the full analytic results here.

For an input state $|\Psi_{e+}\rangle | \alpha_i \rangle$, where
$|\Psi_{e\pm}\rangle = (|00\rangle \pm |11\rangle)/\sqrt{2}$, the
state after a single round of the protocol (i.e. at time $t_4$), is
given by $|\Psi_{e+}\rangle | \alpha_{e} \rangle$, with
\begin{equation}
\alpha_{e} = \beta_{34} + e^{-\kappa (\tau_3+\tau_4)}(\beta_{12} +
e^{-\kappa(\tau_1 + \tau_2)} \alpha_i), \nonumber
\end{equation}
where $\beta_{12} = -(\epsilon_0/\kappa) [1 - e^{-\kappa \tau_2}]$,
 $\beta_{34} = (\epsilon_0/\kappa) [1 - e^{-\kappa \tau_4}]$,
and the time intervals $\tau_i = t_i-t_{i-1}$ correspond to the
length of each step of the evolution.

For an input state $|\Psi_{o+}\rangle | \alpha_i \rangle$, where
$|\Psi_{o\pm}\rangle = (|01\rangle \pm |10\rangle)/\sqrt{2}$, the
state after a single round is
\begin{multline}
\frac{1}{2}\left[ \left(|01\rangle\langle 01 | + |10\rangle\langle
10 | \right) \otimes |\alpha_{o}\rangle\langle\alpha_{o} | + \ldots
\right. \\ \left. \left(e^{-\Gamma_{01}-\Gamma_{12} - \Phi_{12} -
\Gamma_{23} } |01\rangle\langle 10 | \otimes
|\alpha_{o}\rangle\langle \alpha^*_{o} | + \mathrm{h.c.} \right)
\right] \,, \label{eqFinalStateOddParity}
\end{multline}
where
\begin{equation}
\alpha_{o} = \beta_{34} + e^{-i\chi \tau_3}e^{-\kappa
(\tau_3+\tau_4)}(\beta_{12} + e^{-\kappa(\tau_1 + \tau_2)}e^{i\chi
\tau_1} \alpha_i), \nonumber
\end{equation}
and $\Gamma_{01}= \alpha_i^2 [ (1-e^{-2\kappa \tau_1}) +
\frac{\kappa}{\kappa-i\chi} (e^{-2( \kappa - i \chi) \tau_1}  -1) ]
$, $\Gamma_{12}=\alpha_i^2 e^{-2\kappa\tau_1}
(1-e^{-2\kappa\tau_2})(1-e^{2i\chi\tau_1})$, $\Phi_{12}=2i
\mathrm{Im}(\alpha_i\beta_{12}
e^{-\kappa(\tau_2+\tau_1)}e^{i\chi\tau_1})$, and $\Gamma_{23}=
|\alpha_2|^2(1-e^{-2\kappa\tau_3}) + \alpha_2^2
\frac{\kappa}{\kappa-i\chi} (e^{-2( \kappa - i \chi) \tau_1}  -1)$,
with $\alpha_2 = \beta_{12} + e^{-\kappa(\tau_1 + \tau_2)}e^{i\chi
\tau_1} \alpha_i$. The real part of the exponent
${-\Gamma_{01}-\Gamma_{12} - \Phi_{12} - \Gamma_{23} }$ corresponds
to a dephasing error (see below), while the imaginary part leads to
a deterministic phase rotation which is ultimately corrected by
applying the single qubit unitary $U_z$ [see Fig.
\ref{figPhaseSpaceCartoon} (a)], after the homodyne measurement
stage of the protocol.

The timescales $\tau_2$, $\tau_3$ and $\tau_4$ are chosen as
follows. $\tau_2$ and $\tau_3$ can be fixed requiring that, for
symmetry, at time $t_3$, the bus mode associated with the input
state $|\Psi_{o\pm}\rangle$ is in the state $\alpha_3 = -\alpha_i$.
This leads to the approximate values $\tau_2 \approx (2 - \kappa^2
\tau_1^2 - \chi^2 \tau_1^2) \alpha_i/\epsilon_0 $ and $\tau_3
\approx \tau_1(1-2\kappa\tau_1)$. $\tau_4$ is chosen such that
$\alpha_{o} = \alpha_i$, i.e. that the bus mode associated with the
input state $|\Psi_{o\pm}\rangle$ is returned to it's initial state
at the end of each round of the protocol.

Using these values of $\tau_i$, it is possible to obtain approximate
expressions for the total dephasing error at the very end of the
protocol. After $N$ rounds, the state corresponding to an initial
state $|\Psi_{o+}\rangle$ is $\rho_o^{(N)} =
\left[(1-p)|\tilde{\Psi}_{o+}\rangle\langle\tilde{\Psi}_{o+}| + p
|\tilde{\Psi}_{o-}\rangle\langle\tilde{\Psi}_{o-}|\right] \otimes
\vert \alpha_i \rangle \langle \alpha_i \vert$, where $p =
(1-e^{-N\mathrm{Re}(\Gamma_{01}+\Gamma_{12}  + \Gamma_{23})})/2$ is
the probability of a dephasing error, and
$|\tilde{\Psi}_{o\pm}\rangle$ corresponds to $|\Psi_{o\pm}\rangle$,
up to the single qubit correction $U_z$ discussed above. The final
state corresponding to an initial state $|\Psi_{e+}\rangle$ is
$\rho_e^{(N)} = |\Psi_{e+}\rangle \langle \Psi_{e+} | \otimes \vert
\alpha^{(N)}_e \rangle \langle \alpha^{(N)}_e \vert$, where
$\alpha^{(N)}_e \approx \alpha_i + N \alpha_i \chi^2 \tau_1^2$.

Writing the expression for the dephasing error in terms of $\tau_1$
gives the approximate expression $p \approx 4 N \alpha_i^2 \kappa
\chi^2 \tau_1^3 /3 + 4 N \alpha_i^3 \kappa \chi^2 \tau_1^2
/\epsilon_0$. The first term in this expression originates from
decoherence on the `rotation' parts of the evolution (i.e. $t_0 \to
t_1$ and $t_2 \to t_3$), whereas the second term originates from the
first `displacement' part ($t_1 \to t_2$). Note that no decoherence
results from the second displacement stage ($t_3 \to t_4$), as the
bus mode is not in a superposition state during these parts of the
evolution.

\begin{figure}[!hbt]
\includegraphics{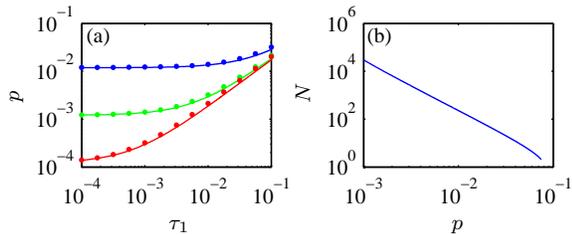}
\caption{(a) Exact (curves) and approximate (points) values of the
final dephasing error against $\tau_1$. Other parameters: $A = 1.5$,
$\alpha_i=2$, $\kappa = 0.05$, $\chi=1$, $\epsilon_0 = 10^2, 10^3,
10^4$, (top to bottom). (b) Overhead, $N$, required to reach a given
error rate, $p$, for $A = 1.5$, $\alpha_i=2$, $\kappa = 0.05$,
$\chi=1$, $\epsilon_0 = 10^5$.} \label{figErrorsOverhead}
\end{figure}

The total number of required rounds of the protocol, $N$, is fixed
by the requirement that $|\alpha^{(N)}_e\rangle$ and
$|\alpha_i\rangle$ are distinguishable by a $X-$homodyne measurement
at the end of the protocol. For coherent states, this implies that
$\alpha^{(N)}_e - \alpha_i = A$, where $A \gtrsim 1$ is a constant
that depends on the required error in distinguishing the states
$|\Psi_{o\pm}\rangle$ from the states $|\Psi_{e\pm}\rangle$. This
requirement gives $N = A / \alpha_0 \chi^2 \tau_1^2$. Thus the total
error is well approximated by
\begin{equation}
p \approx \frac{4 A \alpha_i \kappa \tau_1}{3}   + \frac{4 A
\alpha_i^2 \kappa}{\epsilon_0}\,. \label{eqApproxTotalError}
\end{equation}
The first term in this expression can be made arbitrarily small by
decreasing the value of $\kappa \tau_1$. The second term in this
expression is proportional the ratio $\kappa/\epsilon_0$. Thus
errors can be minimized provided the strength of the classical
driving field can be made large compared to the photon loss rate. An
important point to note is that both terms are independent of the
ratio $\kappa / \chi$, and therefore the scheme is robust to losses
in the bus mode even in the regime $\kappa > \chi$, where the
strength of the qubit-bus interaction is weak compared to the loss
rate. The approximate error, together with the exact values for the
same values of the parameters $\tau_i$, are plotted in Fig.
\ref{figErrorsOverhead}.

\noindent

The price to be paid for this robustness is an increase in the
overhead cost of the scheme: the first term in Eq.
(\ref{eqApproxTotalError}) can be reduced by reducing $\tau_1$, but
as $\tau_1$ is reduced, the required number of rounds of the
protocol required increases. If $p$ is dominated by the first term,
then the required number of rounds is $N \approx (16A^3 \alpha_i /9)
(\kappa/\chi)^2 / p^2$. Thus the overhead cost grows quadratically
with the required reduction in the errors (see Fig.
(\ref{figErrorsOverhead})).

In view of this overhead, it is worth briefly considering the effect
of errors in the scheme. Stochastic noise in the control pulses used
to implement the protocol tend to broaden final state of the bus
mode, making the odd and even parity outcomes harder to distinguish
by the $X-$ homodyne measurement. Treating the noise in each pulse
as independent Gaussian processes, the final uncertainty in the real
part of $\alpha_0^{(N)}$ is found to be $\sigma^2({\alpha_0^{(N)}})
= N \sigma^2({\alpha_0 }) \approx N [\sigma^2(\beta_{12}) +
\sigma^2(\beta_{34}) + \beta^2_{12}\sin^2\theta\sigma^2(\theta)]
\approx 2 N \sigma^2(\beta_{12}) + 4 \alpha_0 A \sigma^2(\theta)$,
where $\sigma^2(\ldots)$ denotes the variance in each parameter.
Thus the noise will not significantly affect the outcome of the
X-homodyne measurement provided $\sigma({\beta_{12}}) \lesssim
N^{-\frac{1}{2}}$ and $ \sigma(\theta) \lesssim (4 \alpha_0
A)^{-\frac{1}{2}}$. If these conditions cannot be met then the
scheme can still be made to work by increasing the number of rounds,
since the peak separation grows linearly in $N$, while the peak
width grows only as $N^{\frac{1}{2}}$. Another potentially important
source of noise is single qubit errors. These can potentially build
up during the implementation of the two-qubit operation, leading to
an effective error rate that scales with $N$. Thus our scheme is
particularly suited to systems in which the single qubit decoherence
rate is small compared to the bus damping rate. Fortunately, there
are many systems (such as trapped ions \cite{Leibfried2003}) where
the single qubit error rate is much longer than all other
timescales.

In conclusion, we have proposed a method for implementing two-qubit
operations via a quantum bus, which is extremely robust to
dissipation of the bus degree of freedom. Our method implements a
parity measurement of the two-qubit system, which, when augmented
with single qubit operations, is sufficient for universal
computation. The residual errors in our scheme can be made
negligible even when the strength of the qubit-bus interaction is
weak compared to the loss rate. We anticipate that this robustness
will permit bus-mediated QIP in a variety of physical setups,
including superconducting systems, trapped ions or atoms, and
all-optical systems.

{\em Acknowledgments}: SDB was supported by the {\sc eu}, the {\sc
cqct}, and the {\sc epsrc}. SDB thanks Aggie Branczyk, Alexei
Gilchrist, Pieter Kok, Bill Munro, Tim Spiller, and Tom Stace for
useful conversations.

\bibliography{LossTolerantBSA}

\end{document}